\documentclass[aps,prl,twocolumn,showpacs,showkeys,footinbib,amsmath,amssymb,superscriptaddress]{revtex4-2}

\usepackage{amsmath}
\usepackage{amssymb}
\usepackage{graphicx}
\usepackage{bm}
\usepackage{color}
\usepackage{relsize}
\usepackage{braket}
\usepackage{bbold}
\usepackage{txfonts}
\usepackage{epstopdf}
\usepackage{multirow}

\PassOptionsToPackage{colorlinks=true,
	linkcolor=blue,
	citecolor=blue,
	urlcolor=blue}{hyperref}
\usepackage{hyperref}

\begin{document}

\title{Symmetry-Driven Unconventional Magnetoelectric Coupling in Perovskite Altermagnets: From Bulk to the Two-Dimensional Limit}

\author{Zhou Cui}
\affiliation{Ningbo Institute of Digital Twin, Eastern Institute of Technology, Ningbo, Zhejiang 315200, China}
\affiliation{Eastern Institute for Advanced Study, Eastern Institute of Technology, Ningbo, Zhejiang 315200, China}

\author{Ziye Zhu}
\email{zyzhu@eitech.edu.cn}
\affiliation{Ningbo Institute of Digital Twin, Eastern Institute of Technology, Ningbo, Zhejiang 315200, China}
\affiliation{Eastern Institute for Advanced Study, Eastern Institute of Technology, Ningbo, Zhejiang 315200, China}

\author{Xunkai Duan}
\affiliation{Ningbo Institute of Digital Twin, Eastern Institute of Technology, Ningbo, Zhejiang 315200, China}
\affiliation{School of Physics and Astronomy, Shanghai Jiao Tong University, Shanghai 200240, China}

\author{Bowen Hao}
\affiliation{Ningbo Institute of Digital Twin, Eastern Institute of Technology, Ningbo, Zhejiang 315200, China}

\author{Xianzhang Chen}
\affiliation{Ningbo Institute of Digital Twin, Eastern Institute of Technology, Ningbo, Zhejiang 315200, China}
\affiliation{Eastern Institute for Advanced Study, Eastern Institute of Technology, Ningbo, Zhejiang 315200, China}

\author{Jiayong Zhang}
\affiliation{Eastern Institute for Advanced Study, Eastern Institute of Technology, Ningbo, Zhejiang 315200, China}
\affiliation{School of Physical Science and Technology, Suzhou University of Science and Technology, Suzhou, 215009, China}

\author{Tong Zhou}
\email{tzhou@eitech.edu.cn}
\affiliation{Ningbo Institute of Digital Twin, Eastern Institute of Technology, Ningbo, Zhejiang 315200, China}
\affiliation{Eastern Institute for Advanced Study, Eastern Institute of Technology, Ningbo, Zhejiang 315200, China}

\date{December 30, 2025}

\begin{abstract}

 The emergence of altermagnets establishes a new paradigm for multiferroics. Unlike conventional multiferroics relying on direct magnetoelectric coupling, multiferroic altermagnets host a crystal-symmetry-mediated magnetoelectric interaction that is intrinsically more efficient and robust. Among candidate material platforms, layered perovskites are particularly appealing owing to their structural diversity and synthetic versatility. However, magnetoelectric properties at the two-dimensional scale remain largely unexplored, hindering their applicability in miniaturized, highly integrated devices. Here, we systematically investigate the dimensional evolution of ferroelectric polarization and magnetism in perovskite systems through symmetry analysis. We demonstrate that altermagnetism can persist in the two-dimensional limit, yet is strongly constrained by the magnetic configuration---with only C-type antiferromagnetic order supporting it. Based on mode-decomposition calculations, we further reveal that symmetry-restricted multimode couplings simultaneously govern ferroelectric polarization and altermagnetic spin splitting. Finally, combined with first-principles calculations, we propose several strategies to lift the magnetic-configuration constraint, extending the range of viable altermagnetic systems. These results underscore the critical role of dimensionality in symmetry-driven magnetoelectric coupling in perovskite altermagnets and pave the way toward next-generation electrically controlled spintronic and multiferroic devices.     

\end{abstract}
\maketitle

Magnetoelectric multiferroics, in which spin polarization can be manipulated by an electric field and vice versa, are highly sought after for applications in high-density, low-energy devices~\cite{Eerenstein2006Nature, Cheong2007NM, Dong2015AP}. This prospect, however, has been limited by the intrinsically weak magnetoelectric coupling in conventional multiferroics, arising from the fundamental dichotomy of $\mathcal{P}$- and $\mathcal{T}$-symmetry breaking in electric and magnetic orders~\cite{Hill2000JPCB, Dong2019NSR}. Recently, unconventional magnets, particularly altermagnets (AM) that unify features of ferromagnets and antiferromagnets, have attracted widespread attention~\cite{Wu2007PRB, Hayami2019JPSP, Yuan2020PRB, Libor2020Crystal, JunWeiLiu2021NC, zhu2024observation, Krempasky2024, zhang2025crystal, jiang2025metallic, SongAMNature2025, Cao2025PRL, wang2025arXiv, Smejkal2022PRX1, Smejkal2022PRX2, Bai2024AFM, Song2025NRM, bhowal2025arxiv, Cheong2025npj, Guo2025AM, Yu2025AM, Fukaya2025JPCM}, while also providing a new research paradigm for multiferroics~\cite{Duan2025PRL, Gu2025PRL, Smejkal2024Arxiv, Zhu2025NanoLetters, Zhu2025SCPMA, SunWei2025Arxiv, Cao2024Arxiv, Bhowal2025PRL, Guo2025Arxiv, UrruPRB2025, Sun2025AM, Sun2025AdvSci, Zhu2025PRL, Zhu2025AMSFET, Zhang2025AS, Chen2025ACSAMI, PengNPJQM2025, Huang2025PRL, Ding2025PRB}. Altermagnetism is characterized by spin-group symmetry and is governed by real-space crystal symmetry~\cite{Spingroup_SongPRX, Spingroup_FangPRX, Spingroup_LiuPRX, Spingroup_LiuMagnon}, a fundamental constraint that is also shared by ferroelectricity. This symmetry compatibility enables a crystal-symmetry-mediated unconventional magnetoelectric coupling, which is inherently more efficient and robust than conventional mechanisms.

Perovskites constitute a large family of compounds with versatile physical properties and are among the most extensively studied materials in condensed matter physics and materials science~\cite{tilley2016perovskites}. In the context of multiferroics, perovskites are widely regarded as some of the most promising candidates for practical applications, exemplified by BiFeO$_3$ and its related compounds~\cite{Yang2015ARMR}. Notably, perovskites also provide a key platform for multiferroic altermagnets, including antiferroelectric altermagnets (AFEAM)~\cite{Duan2025PRL}, e.g., GdFeO$_3$-type BiCrO$_3$~\cite{Duan2025PRL}, and ferroelectric altermagnets (FEAM)~\cite{Zhu2025NanoLetters}, e.g., the Ruddlesden-Popper (RP) phase Ca$_3$Mn$_2$O$_7$~\cite{Gu2025PRL,Libor2024arXiv}. Meanwhile, functional electronic devices are predominantly realized in thin-film or low-dimensional geometries, making the investigation of material properties in the two-dimensional (2D) limit crucial for device miniaturization and high-density integration in the post-Moore era~\cite{Fiori2014NN,Liu2019Nature}. However, existing studies of perovskite multiferroic altermagnets have thus far focused almost exclusively on bulk crystalline phases. This raises the following questions: Can these perovskite materials retain multiferroic altermagnetic order in the 2D limit, and how does crystal symmetry evolve across the dimensional crossover?

In this work, we perform a systematic symmetry analysis of RP phase and GdFeO$_3$-type perovskites to provide an in-depth investigation of dimensionality effects on the ferroelectric and altermagnetic properties at both bulk and 2D scales. We find that these two bulk structures can be converted into the same 2D FEAM perovskites via exfoliation and surface effects; however, the emergence of altermagnetism is constrained by the magnetic configuration, with only the C-type antiferromagnetic order preserved. The ferroelectric polarization originates from the bulk specific alternating in-plane polarization, while the $m_z$ symmetry plays a crucial role in enabling altermagnetism. Furthermore, mode-decomposition calculations for the Ca–Mn–O system reveal that symmetry-restricted multimode couplings simultaneously govern ferroelectric polarization and altermagnetic spin splitting in 2D perovskite multiferroic altermagnets.

\begin{figure*}[htbp]
	\centering
	\includegraphics[width=0.85\textwidth]{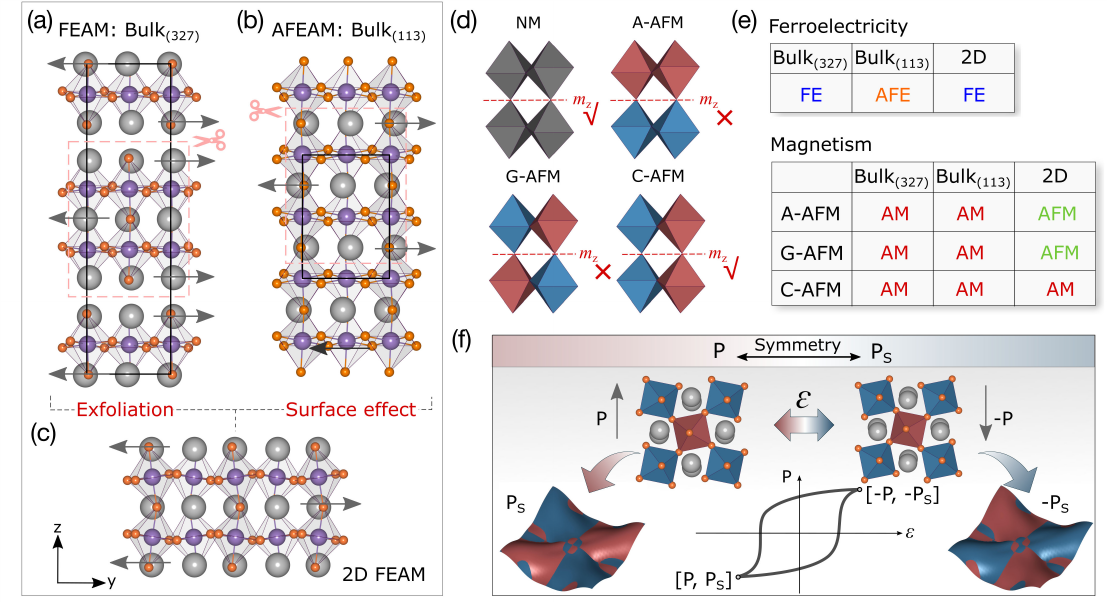}
	\caption{ 
		(a) Crystal structures of bulk FEAM Ruddlesden–Popper perovskites [space group: $\textit{Cmc2}_1$, labeled as Bulk$_{(327)}$] and (b) AFEAM GdFeO$_3$-type perovskites [space group: $Pnma$, labeled as Bulk$_{(113)}$]. Black arrows indicate the direction of ferroelectric polarization, and black solid lines delineate the crystallographic unit cell.
		(c) Two-dimensional slabs derived from the bulk structures via exfoliation or surface-induced dimensional reduction.
		(d) Schematic illustration of the symmetry dependence of magnetic octahedra on the $m_z$ mirror symmetry under different magnetic configurations in 2D perovskites. Red and blue colors denote opposite spin channels.
		(e) Summary of ferroelectric and magnetic orders in bulk and 2D perovskites.
		(f) Schematic illustration of symmetry-driven unconventional magnetoelectric coupling in 2D perovskite altermagnets.
	}
	\label{fig:F1}
\end{figure*}
\begin{table*}[htbp]
	\centering
	\setlength{\tabcolsep}{6pt}
	\caption{ Symmetry operations excluded upon considering magnetic configurations in bulk and 2D perovskites (connecting opposite-spin sublattices). Note that in Bulk$_{(327)}$, the $m_x$ and $C_{2y}$ operations can each follow two distinct translation components.}
	\label{Tab:T1}
	\begin{tabular}{c l l l}
		\hline \hline
		Magnetic Order & $\mathrm{Bulk}_{(327)}$ & $\mathrm{Bulk}_{(113)}$ & 2D \\ 
		\hline
		A-AFM & 
		$\{ C_{2y}t,\ m_{z},\ m_{z}t \}$ & 
		$\{ C_{2y}t,\ C_{2z}t,\ m_{y}t,\ m_{z}t \}$ & 
		$\{ C_{2x}t,\ m_{z}t \}$ \\
		
		C-AFM & 
		$\{ C_{2y}t, \ m_{x}t \}$ & 
		$\{ C_{2x}t,\ C_{2y}t,\ m_{x}t,\ m_{y}t \}$ & 
		$\{ C_{2x}t,\ m_{y}t \}$ \\
		
		G-AFM & 
		$\{ m_{x}t,\ m_{z},\ m_{z}t \}$ & 
		$\{ C_{2x}t,\ C_{2z}t,\ m_{x}t,\ m_{z}t \}$ & 
		$\{ m_{y}t,\ m_{z}t \}$ \\
		\hline \hline
	\end{tabular}
\end{table*}

More intriguingly, we discover that the crystal symmetry can be further engineered to lift the magnetic-configuration constraint in the 2D limit through several strategies, including superlattice engineering, shear strain, electric fields, and substrate engineering. Combined with first-principles calculations, these approaches can convert other magnetic configurations into altermagnets and even realize ferroelectric fully compensated ferrimagnets, thereby extending the range of viable unconventional multiferroics. These results clarify the underlying mechanism of unconventional magnetoelectric coupling in perovskite multiferroic altermagnets and establish a pathway for designing electrically controllable 2D spintronic and multifunctional devices~\cite{Zutic2004RMP,Zutic2019MT, Zhu2025Arxiv,Zhu2025AMSFET,Zhou2021PRL, Zhou2022NC,Konstantin2025PRL,Huang2025arXiv}.

\begin{figure*}[htbp]
	\centering
	\includegraphics[width=0.85\textwidth]{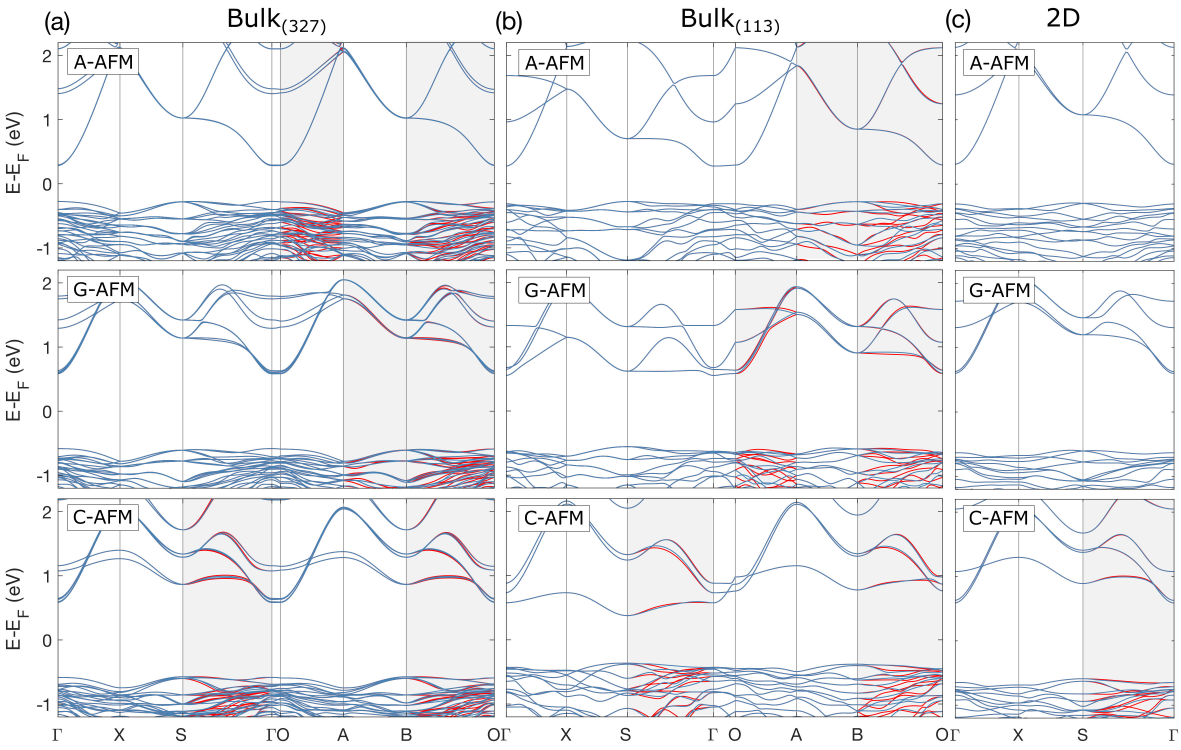}
	\caption{
		Spin-resolved band structures of the Ca-Mn-O multiferroic altermagnets for (a) Bulk$_{(327)}$, (b) Bulk$_{(113)}$, and (c) 2D structure. A-, G-, and C-type AFM orders are considered. Red (blue) lines correspond to spin-up (spin-down) channels. Gray shaded regions highlight momentum paths exhibiting spin splitting. The corresponding first Brillouin zones are provided in Fig.S3.
	}
	\label{fig:F2}
\end{figure*}


Bulk RP phase perovskites with the chemical formula A$_3$B$_2$O$_7$ (hereafter denoted as Bulk$_{(327)}$) crystallize into a van der Waals (vdW) layered structure with the polar space group \textit{Cmc2}$_1$, as shown in Fig.~\ref{fig:F1}(a). Relative to the high-symmetry parent $I4/mmm$ structure, the RP phase is strongly distorted, featuring A-site cation displacements, BO$_6$ octahedral rotations and tilts, and Jahn–Teller distortions. Among these, the A-site displacements primarily generate ferroelectric polarization (see Fig.~S1), while the B-site cations carry the magnetic moments. This symmetry reduction simultaneously satisfies the criteria for altermagnetism, rendering Bulk$_{(327)}$ a switchable FEAM. In contrast, bulk GdFeO$_3$-type perovskites with the general formula ABO$_3$ (denoted as Bulk$_{(113)}$) adopt the \textit{Pnma} space group. Although Bulk$_{(113)}$ exhibits lattice distortions similar to those in Bulk$_{(327)}$, the antiparallel displacements of adjacent A-site layers cancel out, resulting in an AFEAM state, as illustrated in Fig.~\ref{fig:F1}(b). 

Remarkably, both exfoliation from Bulk$_{(327)}$~\cite{zhou2021hybrid,zhou2025out} and surface engineering of Bulk$_{(113)}$~\cite{lu2018unusual} yield the same 2D perovskite structure with the chemical formula A$_6$B$_4$O$_{14}$ (hereafter simply denoted as 2D), as shown in Fig.~\ref{fig:F1}(c). To avoid surface or interface reconstructions associated with the so-called ``polar catastrophe''~\cite{harrison1978polar, lu2018unusual, PhysRevB.108.245304}, we focus on A-site cations with +2 charge, ensuring that each AO layer remains charge neutral. In this well-defined 2D structure, the absence of cancellation between adjacent A-site layers gives rise to a net in-plane ferroelectric polarization, as illustrated in Fig.~\ref{fig:F1}(c) and (e). Indeed, the ferroelectricity in Bulk$_{(113)}$ exhibits a pronounced even-odd layer dependence, as discussed in Fig.~S2.

The magnetic properties exhibit a qualitatively different dimensional dependence. In bulk, both Bulk$_{(327)}$ and Bulk$_{(113)}$ display altermagnetism under A-, G-, and C-AFM orderings, while in the 2D limit only the C-AFM state preserves altermagnetism, as illustrated in Fig.~\ref{fig:F1}(e). Accordingly, in 2D perovskites, the coexistence of ferroelectricity and altermagnetism is restricted to the magnetic configuration.

From a symmetry perspective, altermagnetism requires that the two opposite-spin sublattices are connected by crystallographic rotation ($R$) or mirror ($m$) transformations, possibly combined with (rather than directly connected by) translation ($t$) or inversion ($I$) symmetries~\cite{Smejkal2022PRX1}. In 2D systems, the additional $m_z$ and $C_{2z}$ symmetries are also excluded~\cite{PhysRevB.110.054406}. Bulk perovskites possess eight unitary symmetry operations in their geometric space groups: $\{E, C_{2y}, m_x, m_z\}$ for Bulk$_{(327)}$, each followed by two appropriate lattice translations; and $\{E, I, C_{2x}, C_{2y}, C_{2z}, m_x, m_y, m_z\}$ for Bulk$_{(113)}$. When considering different AFM configurations, half of these operations are excluded, as listed in Tab.~\ref{Tab:T1}. Importantly, the remaining operations do not involve direct $t$ or $I$ symmetries that would forbid altermagnetism, ensuring that all A-, G-, and C-AFM exhibit altermagnetic behavior. For the 2D structure, the geometric space group contains four symmetry operations, $\{E, C_{2y}t, M_x t, m_z\}$, which are also reduced by half when magnetic order is taken into account. For A- and G-type AFM configurations, the remaining operations include the $m_z$ symmetry, enforcing a conventional AFM state. While for the C-AFM configuration, the inherent $m_z$ symmetry is preserved, preventing direct connection between opposite-spin sublattices and thus satisfying the symmetry requirements for altermagnetism, as illustrated in Fig.~\ref{fig:F1}(d) and Tab.~\ref{Tab:T1}.

Next, focusing on the Ca–Mn–O system, we combine symmetry analysis with first-principles electronic-structure calculations to elucidate the dimensional dependence of unconventional magnetoelectric coexistence and coupling in perovskite multiferroic altermagnets. Fig.~\ref{fig:F2} shows the spin-resolved band structures of bulk Ca$_3$Mn$_2$O$_7$, bulk CaMnO$_3$, and 2D Ca$_6$Mn$_4$O$_{14}$ under different AFM configurations. Consistent with the previous analysis, the bulk systems all exhibit momentum-dependent spin splitting, whereas in the 2D limit only the C-type AFM retains altermagnetism. Furthermore, the paths of band splitting vary with the magnetic configuration, as dictated by symmetry. The spin-dependent bands can be described by the Kohn-Sham equation

\begin{equation}
	\left[\frac{1}{2}(\textbf{\emph{k}} - i\nabla)^2 + V_\sigma \right] \psi_\sigma(\textbf{\emph{k}}) = E_\sigma(\textbf{\emph{k}}) \psi_\sigma(\textbf{\emph{k}}),
\end{equation}

\noindent where $\textbf{\emph{k}}$ is the wave vector and $V_\sigma$ the effective potential for spin $\sigma = \uparrow, \downarrow$. An exchange operation $\boldsymbol{O}$ is defined as a symmetry element of the geometric space group that maps spin-up to spin-down, $\boldsymbol{O}V_\uparrow \boldsymbol{O}^{-1} = V_\downarrow$. If $\boldsymbol{O} \textbf{\emph{k}} = \textbf{\emph{k}}$, the spin is degenerate at $\textbf{\emph{k}}$; if instead $\boldsymbol{O} \textbf{\emph{k}} = \textbf{\emph{k'}}$, we have 
$E_\uparrow(\textbf{\emph{k}}) \neq E_\downarrow(\textbf{\emph{k}}) = E_\downarrow(\textbf{\emph{k'}})$, indicating spin splitting with opposite spins at $\textbf{\emph{k}}$ and $\textbf{\emph{k'}}$, characteristic of altermagnetism~\cite{noda2016momentum, Okugawa2018Weakly}.

As an example, for bulk CaMnO$_3$ with C-AFM order, symmetry operations $\boldsymbol{O} \in \{ C_{2x} t, C_{2y} t, m_x t, m_y t \}$, possibly combined with $I$, enforce spin degeneracy along the high-symmetry paths $\Gamma$--$X$--$S$ and $\Gamma$--$O$--$A$--$B$, whereas along $S$--$\Gamma$ and $B$--$O$, no $\boldsymbol{O}$ maps $\textbf{\emph{k}}$ onto itself; instead, $\textbf{\emph{k}}$ is mapped to $\textbf{\emph{k'}}$ along $S'$--$\Gamma$ and $B'$--$O$, resulting in altermagnetic spin splitting. The specific high-symmetry points are labeled in Fig.~S3. For A- and G-type AFM, different sets of $\boldsymbol{O}$ alter the spin-splitting paths accordingly. Similarly, for a 2D slab with C-AFM order, $\boldsymbol{O} \in \{ C_{2y} t, m_x t \}$ maps the opposite-spin sublattices, producing spin splitting along $\Gamma$--$S$, while bands remain spin-degenerate along $\Gamma$--$X$--$S$. In contrast, for A- and G-AFM, all $\textbf{\emph{k}}$ points are mapped onto themselves by $\boldsymbol{O}$, resulting in conventional AFM.

\begin{figure}[!t]
	\centering
	\includegraphics[width=0.48\textwidth]{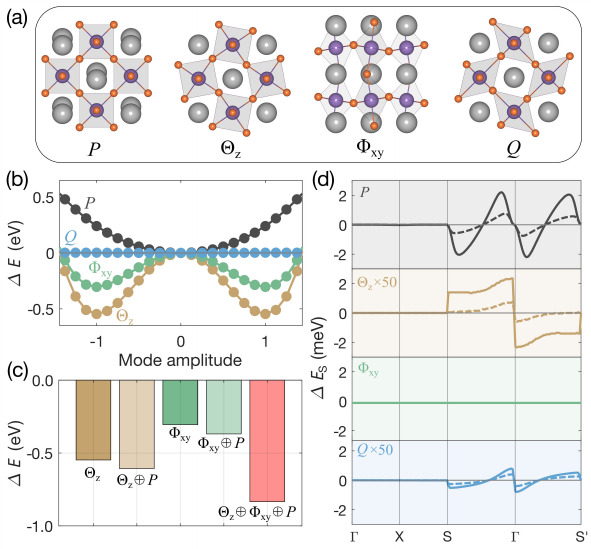}
	\caption{
		(a) Major structural distortions of the ground state in the studied 2D perovskites, including in-plane ferroelectric polarization $P$, in-plane octahedral rotation $\Theta_z$, octahedral tilting $\Phi_{xy}$, and Jahn–Teller distortion $Q$.
		(b) Energy surface as a function of the amplitude of each mode for the 2D high symmetry parent $P4/mmm$ phase in the Ca-Mn-O system.
		(c) Energy gain associated with different lattice distortion modes acting on the $P4/mmm$ phase.
		(d) Altermagnetic spin splitting as a function of the amplitude of different distortion modes; dashed and solid lines correspond to mode amplitudes of 0.5 and 1 (DFT-relaxed structures), respectively.
	}
	\label{fig:F3}
\end{figure}

To gain more insight into the microscopic origin of the magnetoelectric effect in 2D perovskite multiferroic altermagnets, we analyze the different lattice distortion modes. Through mode decomposition, we identify four major atomic distortions: in-plane polarization distortion $P$, in-phase oxygen octahedral rotations $\Theta_z$, out-of-plane oxygen octahedral tilts $\Phi_{xy}$, and the Jahn-Teller lattice distortion $Q$, as illustrated in Fig.~\ref{fig:F3}(a). We project the contribution of each mode onto the ground state structure and calculate the energy surface around the high-symmetry parent $P4/mmm$ reference structure in the Ca--Mn--O system. Fig.~\ref{fig:F3}(b) shows the total energy as a function of the amplitude for each individual distortion. Relatively large energy gains can be seen within characteristic double-well potentials for the $\Theta_z$ and $\Phi_{xy}$ distortions, whereas the in-plane polarization $P$ remains stable. Fig.~\ref{fig:F3}(c) further demonstrates that the collective $\Theta_z \oplus \Phi_{xy} \oplus P$ combination strongly lowers the total energy, indicating that ferroelectricity in this system arises from a multimode coupling effect, analogous to hybrid improper ferroelectricity~\cite{benedek2011hybrid, Chen2019PRL}.

Similarly, we evaluate the contribution of each distortion mode to the spin splitting, as shown in Fig.~\ref{fig:F3}(d). We find that the $P$, $\Theta_z$, and $Q$ modes all contribute positively to the altermagnetic spin splitting, such that larger distortion amplitudes lead to larger spin splittings, whereas the $\Phi_{xy}$ mode does not induce any splitting due to symmetry constraints. Notably, the spin splitting generated by the three modes individually, when summed ($\sim$2 meV), remains much smaller than that of the ground state ($\sim$20 meV), indicating that altermagnetic spin splitting in the 2D Ca--Mn--O structure originates from the cooperative contribution of multiple distortion modes. Furthermore, spin-resolved band structures for different $B$-site elements are calculated, and the distortion amplitudes of the various modes are quantitatively analyzed (see Fig.~S4). These results further demonstrate a strong correlation between between the magnitude of altermagnetic spin splitting and the multimode coupling effect.
  
At this stage, we have clarified the dimensional dependence and microscopic origin of the magnetoelectric coupling in perovskite multiferroic altermagnets, with the key finding that altermagnetism can persist in the 2D limit but is strongly constrained by the magnetic configuration, such that only the C-AFM order is supported. Through A- and B-site elemental substitution calculations, we further find that many 2D perovskite materials do not stabilize the C-AFM order, as shown in Fig.~S5. This raises the question: can this magnetic-configuration constraint be lifted?

\begin{figure}[!t]
	\centering
	\includegraphics[width=0.48\textwidth]{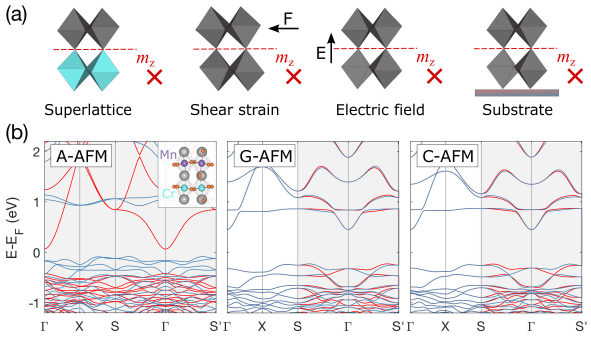}
	\caption{
		(a) Several strategies to break $m_z$ symmetry in real space, including superlattice engineering, shear strain, electric field, and substrate engineering. (b) Spin-resolved band structures of different magnetic orders in the 2D Ca-Mn/Cr-O superlattice (chemical formula Ca$_6$Mn$_2$Cr$_2$O$_{14}$).
	}
	\label{fig:F4}
\end{figure}

In general, there are two rules to design altermagnetism: (i) breaking specific symmetries without considering the magnetic order; and (ii) preserving specific symmetries after the magnetic configuration is established, thereby preventing them from serving to connect opposite-spin sublattices. These specific symmetries include $t$ and $I$ (and $m_z$, $C_{2z}$ in the 2D case). In both cases, the $R$-related rotational symmetries must still be preserved. As discussed in Fig.~\ref{fig:F1}(d) and Tab.~\ref{Tab:T1}, the emergence of C-AFM as the 2D altermagnetic state arises precisely from the second mechanism: the inherent $m_z$ symmetry remains intact, preventing it from serving as the symmetry operation that connects opposite-spin sublattices. Once the magnetic configuration is identified as a non–C-AFM state, the second mechanism is no longer applicable, and $m_z$ symmetry must be broken in the crystal space to realize altermagnetism. Here, we propose four general strategies to achieve this, including superlattice engineering, shear strain, electric fields, and substrate engineering, as illustrated in Fig.~\ref{fig:F4}(a).

In Fig.~\ref{fig:F4}(b) and S6, we show the band structures obtained after applying the aforementioned strategies. We find that, upon breaking the $m_z$ symmetry, G-AFM order in 2D perovskites transforms from a conventional AFM into AM. Interestingly, under the reduced symmetry, A-AFM order exhibits the behavior of a fully compensated ferrimagnet—zero net magnetization but spin splitting across the entire Brillouin zone, as the opposite-spin sublattices are no longer connected by any symmetry operation. To our knowledge, this represents the first report of a 2D ferroelectric fully compensated ferrimagnet.

In summary, based on symmetry analysis and first-principles calculations, we have systematically explored the dimensional evolution and microscopic origin of unconventional magnetoelectric coupling in perovskite multiferroic altermagnets. We further propose a series of potential strategies to realize 2D multiferroic altermagnets, including fully compensated multiferroic ferrimagnets. Notably, in the 2D perovskite multiferroic altermagnets considered here, an applied electric field can reverse the spin splitting, as illustrated in Fig.~\ref{fig:F1}(d). Such robust, electric-field–controllable spins in the 2D limit can be readily integrated into van der Waals heterostructures, providing a versatile platform for emerging electrically controlled spin phenomena~\cite{Zutic2004RMP,Zutic2019MT, Zhu2025Arxiv,Zhu2025AMSFET,Zhou2021PRL, Zhou2022NC,Konstantin2025PRL}. Moreover, by leveraging the intrinsic advantages of altermagnets, multiferroic altermagnets can substantially outperform conventional multiferroics in terms of spin dynamics, read/write speed, and high-density integration, offering promising opportunities for next-generation memory technologies.

This work is supported by the National Natural Science Foundation of China (12474155, 12447163, 12504108, and 11904250), the Zhejiang Provincial Natural Science Foundation of China (LR25A040001), the China Postdoctoral Science Foundation (2025M773440), and the U.S. DOE, Office of Science BES, Award No. DE-SC0004890 (I.\v{Z}.). The computational resources for this research were provided by the High Performance Computing Platform at the Eastern Institute of Technology, Ningbo.

\medskip

\bibliography{reference}

\end{document}